\documentclass[amssymb,showpacs,aps]{revtex4}
\usepackage{epsfig}
\begin{document}
\title{Linearized solutions of the Einstein
equations within a Bondi-Sachs framework, and implications
for boundary conditions in numerical simulations}

\author{Nigel T. Bishop}
\affiliation{
Department of Mathematical Sciences,
University of South Africa, P.O. Box 392, Unisa 0003,
South Africa}

\date{14 March 2005}

\begin{abstract}
We linearize the Einstein equations when the metric is
Bondi-Sachs, when the background is Schwarzschild or
Minkowski, and when there is a matter source in the form
of a thin shell whose density varies with time and angular
position. By performing an eigenfunction decomposition, we
reduce the problem to a system of linear ordinary differential
equations which we are able to solve. The solutions are
relevant to the characteristic formulation of numerical
relativity: (a) as exact solutions against which computations
of gravitational radiation can be compared; and (b) in
formulating boundary conditions on the $r=2M$ Schwarzschild
horizon.
\end{abstract}

\pacs{04.25.Nx, 04.25.Dm, 04.30.Db}

\maketitle

\section{Introduction}
\label{s-in}
This work is motivated by the needs of numerical relativity,
specifically the characteristic, or null
cone, formulation used in the PITT code~\cite{hpn,roberto} (and see
also~\cite{cce,mat,particle,fission}). It should be noted that the
PITT code is not
the only approach to characteristic numerical relativity --
for example, see~\cite{ntb90,rdi1,rdi2,philiptoni,siebel1,siebel2}.
The PITT code exhibits long-term stability
in evolving single black hole spacetimes~\cite{wobble}, and has also
been successfully applied to the problem of
nonlinear scattering of gravitational radiation
by a black hole~\cite{mod}. However, it has not been possible to compute
gravitational radiation emitted in astrophysically interesting
scenarios involving matter -- for example, a star in close orbit
around a black hole~\cite{particle}. Variations of the method
of calculating the gravitational radiation have been
investigated~\cite{newnews} and~\cite{mod,zlo}, but the problem has
not been resolved. An important difficulty in trying to
analyze the problem is that, while
methods such as the quadrupole formula indicate approximately what the
final result of the computation should be, there is no guidance
concerning the intermediate quantities. Thus, in this paper we
construct analytic solutions containing gravitational radiation
and matter, against which a numerical code can be tested.

In addition, we discover an unexpected issue concerning the boundary
condition commonly imposed on a black hole horizon -- the details
are given in Sec.~\ref{s-d}.

Linearization and eigenfunction decomposition
are tools commonly used to obtain an analytic understanding
of mathematically complex systems. The Einstein equations are a
particularly complicated system of nonlinear partial differential
equations, and there is an extensive literature on linearization
techniques. Work on linearization within the characteristic
formalism includes the Newonian correspondence method~\cite{newt1,
newt2,iww}, which is related to linearization about a Minkowski
background. More recently~\cite{man,hus}, the Teukolsky
equation~\cite{teuk} has been used as a starting point to
investigate gravitational radiation on a Schwarzschild background;
but we do not use that formalism here
because the coordinates are not those of the PITT code.
The key point about the approach presented here is that we
use the same coordinates and metric
variables as in the code, so that comparisons
between numerical and analytic solutions can easily be made.
We also allow for the presence of a matter source.

We seek a physical situation that is without symmetry and that
radiates gravitationally, but in which
the linearized problem has a solution in simple analytic form.
The model problem should be relevant to the physical problem of a star
in quasi-circular orbit around a black hole.
We achieve these goals by considering a Schwarzschild black hole of mass
$M$ (with the case $M=0$ corresponding to a Minkowski background),
surrounded by a thin spherical shell of matter
whose density varies in time $u$ as $e^{i\nu u}$
(with the static case recovered by setting $\nu=0$), and whose angular
dependence is that of a spherical harmonic.
We find a solution for all metric variables in simple
analytic form, except when $M\ne 0$ and $\nu\ne 0$
in which case solutions are obtained as power series.

The calculations reported in this paper are very long, and
would have been impossible to perform without the use of a
computer algebra system, in this case Maple.

In Sec.~\ref{s-f} we give a summary of the formalism concerning
the Bondi-Sachs metric in numerical relativity, together with the
formalism of spin-weighted spherical harmonics. Next, in
Sec.~\ref{s-si}, we specify the model problem to be solved and work
out the resulting Einstein equations, reducing the problem to
coupled linear ordinary differential equations. Solutions are found
in Sec.~\ref{s-so}, and presented according to the values
of the parameters $\nu$ and $M$. The paper concludes, in Sec.~\ref{s-c},
with a discussion of appropriate boundary conditions at a black hole
horizon, and an outline of opportunities for taking this work further.

\section{Formalism}
\label{s-f}
\subsection{Bondi-Sachs metric}

The formalism for the numerical evolution of Einstein's equations, in
null cone  coordinates, is well known~\cite{hpn,roberto,cce,mat,rai83,
bondi}. For the sake of completeness, 
we give a summary of those aspects of the formalism that will be used here.
We start with coordinates based upon a family of outgoing null
hypersurfaces.
We let $u$ label these hypersurfaces, $x^A$ $(A=2,3)$, label
the null rays and $r$ be a surface area coordinate. In the resulting
$x^\alpha=(u,r,x^A)$ coordinates, the metric takes the Bondi-Sachs
form~\cite{bondi,sachs}
\begin{eqnarray}
 ds^2  =  -\left(e^{2\beta}(1 + {W \over r}) -r^2h_{AB}U^AU^B\right)du^2
\nonumber \\
        - 2e^{2\beta}dudr -2r^2 h_{AB}U^Bdudx^A 
        +  r^2h_{AB}dx^Adx^B,
\label{eq:bmet}
\end{eqnarray}
where $h^{AB}h_{BC}=\delta^A_C$ and
$det(h_{AB})=det(q_{AB})$, with $q_{AB}$ a unit sphere metric.
We work in stereographic coordinates $x^A=(q,p)$ for which the unit sphere
metric is
\begin{equation}
q_{AB} dx^A dx^B = \frac{4}{P^2}(dq^2+dp^2),
\end{equation}
where
\begin{equation}
        P=1+q^2+p^2.
\end{equation}
We also introduce a complex dyad $q_A$ defined by
\begin{equation}
      q^A=\frac{P}{2}(1,i), \;\;q_A=\frac{2}{P}(1,i)
\end{equation}
with $i=\sqrt{-1}$. For an arbitrary Bondi-Sachs metric,
$h_{AB}$ can then be represented by its dyad component
\begin{equation}
J=h_{AB}q^Aq^B/2,
\end{equation}
with the spherically symmetric case characterized by $J=0$.
We introduce the spin-weighted field
\begin{equation}
U=U^Aq_A,
\end{equation}
as well as the (complex differential) eth operators $\eth$ and $\bar \eth$
(see~\cite{eth} for full details).

Einstein's equations $R_{\alpha\beta}=8\pi(T_{\alpha\beta}
-\frac{1}{2}g_{\alpha\beta}T)$ are classified as: hypersurface
equations -- $R_{11},q^AR_{1A},h^{AB}R_{AB}$ -- forming a hierarchical
set for $\beta,U$ and $W$; evolution equation $q^Aq^B R_{AB}$ for
$J$; and constraints $R_{0\alpha}$. An evolution problem is normally
formulated in the region of spacetime between a timelike or null
worldtube $\Gamma$ and future null infinity, with (free) initial data
$J$ given on $u=0$. Boundary data for $\beta,U,W,J$ that satisfies
the constraints, is required on $\Gamma$. In the case that the
spacetime contains a single Schwarzschild black hole, $\Gamma$ is
usually taken to be the (past) event horizon $r=2M$.

In a general situation the various metric components do not admit
simple physical interpretations, but nevertheless a review of some
particular cases does illustrate the roles of the various
coefficients. Minkowski spacetime is recovered by setting all the
coefficients to zero, i.e. $\beta=W=U=J=0$. Schwarzschild spacetime
has $J=U=0$, and is usually described with $\beta=0$, $W=-2M$, but
can also be represented by $\beta=\beta_c$ (constant) and
$W=(e^{2\beta_c}-1)r-2M$. The quantities $J$ and $U$ are zero when
there is spherical symmetry, and so can be regarded as a measure
of the deviation from spherical symmetry: the quantities are
interlinked, and in the linearized regime contain all the dynamic
content of the gravitational field.

\subsection{Spin-weighted spherical harmonics}
We will be using spin-weighted spherical harmonics~\cite{newp,
golm} using the formalism described in~\cite{mod}. It will
prove convenient to use ${}_s Z_{\ell m}$ rather than the usual
${}_s Y_{\ell m}$ as basis functions, where
\begin{eqnarray}
{}_s Z_{\ell m} &=& \frac{1}{\sqrt{2}} \left[{}_s Y_{\ell m}
   +(-1)^m {}_s Y_{\ell -m}\right] \mbox{ for } m>0 \nonumber \\
{}_s Z_{\ell m} &=& \frac{i}{\sqrt{2}} \left[(-1)^m{}_s Y_{\ell m} 
   -{}_s Y_{\ell -m} \right]\mbox{ for }  m<0 \nonumber \\
{}_s Z_{\ell 0} &=& {}_s Y_{\ell 0},
\end{eqnarray}
and note that~\cite{mod} uses the notation ${}_s R_{\ell m}$
rather than the ${}_s Z_{\ell m}$ used here; we
use a different notation to avoid any confusion with the Ricci
tensor. In the case $s=0$, the $s$ will be omitted, i.e.
$Z_{\ell m}={}_0 Z_{\ell m}$.
Some properties of the ${}_s Z_{\ell m}$ that will be used later are
\begin{itemize}
\item They are orthonormal
\item They are {\bf real}, which is the crucial point -- quantities
like $\bar{\eth}U + \eth\bar{U}$ do not introduce $m$ and $-m$
mode mixing.
\item Defining
\begin{equation}
L_2=-\ell(\ell+1)
\end{equation}
then
\begin{equation}
\bar{\eth}\eth Z_{\ell m} = \eth \bar{\eth} Z_{\ell m}
= L_2 Z_{\ell m}, \;\;
\bar{\eth}\eth^2 Z_{\ell m} = (L_2+2) \eth Z_{\ell m}, \;\;
\bar{\eth}^2\eth^2 Z_{\ell m} = \eth^2 \bar{\eth}^2 Z_{\ell m} 
= L_2(L_2+2) Z_{\ell m}.
\end{equation}
\item The effect of the $\eth$ operator acting on $Z_{\ell m}$ is
\begin{equation}
\eth Z_{\ell m}=\sqrt{-L_2}\;{}_1Z_{\ell m}, \;\;\;
\eth^2 Z_{\ell m}=\sqrt{-(\ell -1)L_2(\ell+2)}\;{}_2Z_{\ell m},
\end{equation}
\end{itemize}

\section{Simplification of the Einstein equations for a
Bondi-Sachs metric}
\label{s-si}

\subsection{The linearized Einstein equations}
\label{s-lin}

Physically, we consider a matter distribution in the form of a
thin, low-density, shell around either a Schwarzschild
black hole or empty space. The shell is spherical at $r=$constant,
but the density $\rho$ can vary around the shell so the problem is not
spherically symmetric. We
regard the density and metric quantities as being small, i.e.
\begin{equation}
\rho, J, \beta, U, w = {\mathcal O}(\epsilon)
\mbox{ where } W=-2M+w,
\end{equation}
with $M$ the mass of the Schwarzschild black hole (and, of course,
the Minkowski case is recovered simply by setting $M=0$).
All terms in the Einstein equations of order
${\mathcal O}(\epsilon^2)$ will be set to zero. That is, we
perform a standard linearization of the Einstein equations.

Using the above ansatz, we found the following forms for the
hypersurface equations and the evolution equation
\begin{equation}
R_{11}: \;\; \frac{4}{r}\beta_{,r}=8 \pi T_{11}
\label{e-b}
\end{equation}
\begin{equation}
q^A R_{1A}: \;\; \frac{1}{2r} \left(
4 \eth \beta - 2 r \eth \beta_{,r} + r \bar{\eth} J_{,r}
+r^3 U_{,rr} +4 r^2 U_{,r} \right) = 8 \pi q^A T_{1A}
\label{e-rq}
\end{equation}
\begin{equation}
h^{AB} R_{AB}: \;\;
(4-2\eth \bar{\eth}) \beta +\frac{1}{2}(\bar{\eth}^2 J + \eth^2\bar{J})
+\frac{1}{2r^2}(r^4\eth\bar{U}+r^4\bar{\eth}U)_{,r} -2 w_{,r}
=8 \pi (h^{AB}T_{AB}-r^2 T)
\label{e-rw}
\end{equation}
\begin{equation}
q^A q^B R_{AB}: \;\;
  -2\eth^2\beta + (r^2 \eth U)_{,r} - 2(r - M) J_{,r}
 - \left( 1 - \frac{2M}{r} \right) r^2 J_{,rr} 
  +2 r (rJ)_{,ur}= 8 \pi q^A q^B T_{AB}.
\label{e-ev}
\end{equation}
In addition, we evaluated the constraint equations; these will be
needed only off the matter shell, and the formulas below are for
the vacuum case
\begin{equation}
R_{00}:\;\;
 \frac{1}{2r^3} \bigg( r(r-2M) w_{,rr}+\eth\bar{\eth} w 
  +2(r-2M) \eth\bar{\eth} \beta
  - M r (\eth \bar{U} + \bar{\eth}U)
-4 r (r-2M) \beta_{,u} - r^3 (\eth \bar{U} + \bar{\eth}U)_{,u}+2 r w_{,u}
 \bigg) = 0
\end{equation}
\begin{equation}
R_{01}:\;\;
 \frac{1}{4r^2} \bigg(2 r w_{,rr}+4 \eth\bar{\eth}\beta
         -(r^2\eth\bar{U}+r^2\bar{\eth}U)_{,r}\bigg)=0
\end{equation}
\begin{equation}
q^A R_{0A}:\;\;
 \frac{1}{4r^2} \bigg( 2r \eth w_{,r}-2 \eth w+ 2 r^2(r-2M)(4 U_{,r}
         + r U_{,rr})+4 r^2 U +r^2(\eth\bar{\eth}U-\eth^2\bar{U})
         +2 r^2 \bar{\eth}J_{,u}-2 r^4  U_{,ur}-4 r^2 \eth\beta_{,u}
         \bigg)=0.
\end{equation}

\subsection{Eigenfunction decomposition}
\label{s-sle}
We first suppose that the various metric quantities can be written as
\begin{equation}
J=J_0(r)\cos(\nu u) \eth^2 Z_{\ell m}, \;\;
U=U_0(r)\cos(\nu u) \eth Z_{\ell m},
\;\;\beta=\beta_0(r)\cos(\nu u) Z_{\ell m},
\;\; w=w_0(r)\cos(\nu u) Z_{\ell m}.
\label{e-an}
\end{equation}
Of course, a general spacetime will not satisfy Eq.~(\ref{e-an}),
but it can be represented by summing over $\ell$ and $m$ and
integrating over $r_0$ and $\nu$. However, for our purposes we
regard $\ell$, $m$, $r_0$ and $\nu$ as fixed.
Later, we will express the time variation in the more usual way as
$e^{i\nu u}$, but for the time being we need the dependence to be
explicitly real. The reason is that we are now using imaginary
quantities as a representation on $S^2$, and this is not related
to the phase.

Having chosen basis functions that are eigenfunctions of the relevant
operators, the result of applying ansatz Eq.~(\ref{e-an}) is that we
reduce Eqs.~(\ref{e-b}) to (\ref{e-ev}) to ordinary differential
equations
\begin{equation}
  \frac{4}{r}\beta_{0,r} \cos(\nu u) Z_{\ell m}=8 \pi T_{11}
\label{e-b1}
\end{equation}
\begin{equation}
\frac{1}{2r} 
\left(4\beta_0-2r\beta_{0,r}+r^3 U_{0,rr} +4r^2 U_{0,r} 
  + (2+L_2)r J_0\right)
\cos(\nu u) \eth Z_{\ell m} =8 \pi q^A T_{1A}
\label{e-q1}
\end{equation}
\begin{equation}
\left(2(2-L_2) \beta_0 +L_2(L_2+2)J_0
+\frac{1}{r^2}(r^4 L_2 U_0)_{,r} -2 w_{0,r}
\right) \cos(\nu u) Z_{\ell m}
=8 \pi (h^{AB}T_{AB}-r^2 T)
\label{e-w1}
\end{equation}
\begin{equation}
\left(-2\beta_0+2U_0 r+r^2 U_{0,r}-2(r - M) J_{0,r} 
  -r^2\left( 1-\frac{2M}{r}\right) J_{0,rr} 
+ 2 r \Psi (J_0+r J_{0,r}) \right)
\cos(\nu u) \eth^2 Z_{\ell m} =8 \pi q^A q^B T_{AB}.
\label{e-j1}
\end{equation}
where $\Psi$ stands for
\begin{equation}
\Psi=-\nu\frac{\sin(\nu u)}{\cos(\nu u)}.
\end{equation}
The functions $\beta_0 (r)$, $U_0 (r)$, $w_0 (r)$ and $J_0 (r)$
can be taken as real in the above equations. Thus, we can now
replace $\cos (\nu u)$ by the real part of $e^{i\nu u}$, and
$\Psi \cos (\nu u)$ by the real part of $-i \nu e^{i\nu u}$.
Any imaginary component introduced into any of
$\beta_0 (r)$, $U_0 (r)$, $w_0 (r)$ or $J_0 (r)$ can now
unambiguously be interpreted as a phase variation.

\subsection{The news}
Linearizing Eq. (B5) in~\cite{hpn}, we find that the gravitational
news is
\begin{equation}
N=\lim_{r \rightarrow \infty}
\left(-\frac{1}{2} r^2 J_{,ru} + \frac{1}{2}\eth^2 \omega 
+\eth^2 \beta \right)
\label{e-ome0}
\end{equation}
where $\omega$ satisfies (see Eqs. (11), (31) and (B1) of~\cite{hpn})
\begin{equation}
2(\omega^2 +\bar{\eth}\eth \log \omega)
=2+\frac{1}{2}(\bar{\eth}^2 J + \eth^2 \bar{J}).
\label{e-ome}
\end{equation}
Now applying the same type of ansatz to $\omega$ as to the other metric
variables, i.e.,
\begin{equation}
\omega=1+\omega_0 Z_{\ell m} e^{i\nu u}
\end{equation}
where $\omega_0={\mathcal O}(\epsilon)$, we find that Eq.~(\ref{e-ome})
simplifies to
\begin{equation}
(4\omega_0+2L_2 \omega_0-L_2(L_2+2)J_{0\infty})
Z_{\ell m} e^{i\nu u}=0,
\label{e-ome1}
\end{equation}
where $J_{0\infty}$ means
\begin{equation}
J_{0\infty}=\lim_{r \rightarrow \infty} J_0(r).
\end{equation}
Solving Eq.~(\ref{e-ome1}) for $\omega_0$, we find
\begin{equation}
\omega_0=\frac{L_2}{2}J_{0\infty}
\end{equation}
so that Eq.~(\ref{e-ome0}) becomes
\begin{equation}
N=\left(-\frac{i\nu}{2} \lim_{r \rightarrow \infty}
\left(r^2 J_{0,r}\right)
 + \frac{L_2}{4}J_{0\infty}
 +\beta_0 \right) \sqrt{-(\ell-1)L_2(\ell+2)}{}_2Z_{\ell m} e^{i\nu u}.
\label{e-N}
\end{equation}

\subsection{Solution procedure}
Eqs.~(\ref{e-b1}) to (\ref{e-j1}) may be solved in the following
way. First, Eq.~(\ref{e-b1}) shows that $\beta_0$ takes a constant
value, say $\beta_{0+}$, in $r>r_0$, and a different constant value,
say $\beta_{0-}$, in $r<r_0$ (although we will set
$\beta_{0-}=0$). Then Eqs.~(\ref{e-q1}) and (\ref{e-j1})
constitute a coupled system of ordinary differential equations
in $U_0(r)$ and $J_0(r)$, and once they have been solved
Eq.~(\ref{e-w1}) can be solved for $w_0(r)$. The computer
algebra does solve the systems for general $\ell$, but some of the
expressions get very complicated and in order to simplify the
presentation we now specialize to the case 
\begin{equation}
\ell=2
\end{equation}
so that $L_2=-6$. In $r>r_0$, we remove all common factors from
Eqs.~(\ref{e-q1}) and (\ref{e-j1}) and they simplify to (with the form
in $r<r_0$ obtained by replacing $\beta_{0+}$ by zero, as well
as $J_{0+}$ by $J_{0-}$ etc.)
\begin{eqnarray}
  4\beta_{0+}+r^3 U_{0+,rr} +4r^2 U_{0+,r} -4 r J_{0+} =0
\label{e-jqu} \\
-2\beta_{0+}+2U_{0+} r+r^2 U_{0+,r}-2(r - M) J_{0+,r} 
  -r^2\left( 1-\frac{2M}{r}\right) J_{0+,rr} 
+ 2 r i\nu (J_{0+}+r J_{0+,r}) =0 \label{e-jqj}
\\
16 \beta_{0+} + 24J_{0+}
-\frac{6}{r^2}(r^4 U_{0+})_{,r} -2 w_{0+,r} =0.
\label{e-jqw}
\end{eqnarray}
The constraints simplify to
\begin{eqnarray}
R_{00}:\;\;
 \frac{1}{2r^3} \bigg( (r^2-2Mr) w_{0+,rr}-6 w_{0+} -12(r-2M) \beta_{0+}
           +12 M r U_{0+} \nonumber \\
-4 r (r-2M)i\nu \beta_{0+} +12 r^3 i \nu U_{0+}+2 r i\nu w_{0+}
 \bigg) = 0 \\
R_{01}:\;\;
 \frac{1}{2r^2} \bigg(r w_{0+,rr}-12 \beta_{0+}
         +6r^2 U_{0+,r}+12 r U_{0+}\bigg)=0 \\
q^A R_{0A}:\;\;
 \frac{1}{2r^2} \bigg(r w_{0+,r}-w_{0+}+4 r^3 U_{0+,r}
         +r^4 U_{0+,rr}+2 r^2 U_{0+}
         -2 M r^3 U_{0+,rr}-8 M r^2 U_{0+,r} \nonumber \\
         -4r^2 i \nu J_{0+}-r^4 i \nu U_{0+,r}-2 r^2 i \nu \beta_{0+}
         \bigg)=0.
\end{eqnarray}

\section{Solutions}
\label{s-so}
Eqs. (\ref{e-jqu}) and (\ref{e-jqj}) can be manipulated to give
\begin{equation}
-2 J_2( 2x +8Mx^2 +i\nu) +2\frac{dJ_2}{dx}\left(2x^2+i\nu x
-7x^3 M\right) +x^3(1-2xM)\frac{d^2J_2}{dx^2}=0
\label{e-S}
\end{equation}
where $J_{2}(x)\equiv d^2J_{0+}/dx^2$ and $x=1/r$. Clearly, two
solutions for $J_{0+}(r)$ are of the form $C_1+C_2/r$, and the
remaining two solutions depend on the values of $\nu$ and $M$.
Once a general solution of Eq.~(\ref{e-S}) has been found, it is
straightforward to calculate $U_{0+}(r)$ and $w_{0+}(r)$. We will
present the solutions in terms of the four cases
$\nu =,\ne 0;M =,\ne 0$.

\subsection{Case $M=\nu=0$ -- Static shell on a Minkowski background}
\label{s-M0nu0}
\begin{eqnarray}
J_{0+} &=& C_{1+} +C_{2+} r^2+ \frac{C_{3+}}{r} +\frac{C_{4+}}{r^3}
\nonumber \\
U_{0+} &=& \frac{2\beta_{0+}}{r} + 2 C_{2+} r+ \frac{2 C_{3+}}{r^2}
-\frac{3 C_{4+}}{r^4} \nonumber \\
w_{0+} &=& -10 \beta_{0+} r+ 12 C_{1+} r -6 C_{2+} r^3
-\frac{6 C_{4+}}{r^2} + C_{5+}
\end{eqnarray}
where the $C_{i+}$, $i=1 \cdots 5$, are constants of integration. An
equivalent solution, with constants of integration
$C_{i-}$, $i=1 \cdots 5$ and with $\beta_{0+}$ replaced by zero, is
applicable in the region $r<r_0$. When we substitute the above
solution into the constraints, we find that $R_{01}=0$ is satisfied
identically, and that, after removing common factors, the other
constraints reduce to
\begin{equation}
R_{00}:\;\; 4r(2\beta_{0+}-3C_{1+})-C_{5+}=0,\;\;\;
q^A R_{0A}:\;\; C_{5+}=0.
\label{e-cM}
\end{equation}
We require that the solution be regular at null infinity, which
means that $J$ must be bounded as $r \rightarrow \infty$, so that
\begin{equation}
C_{2+}=0.
\label{e-infM}
\end{equation}
The constraints lead to $C_{1+}=\frac{2}{3}\beta_{0+}$, and
since $C_{1+}$ is just $J_{0\infty}$, this leads directly to the
news $N=0$, as expected. In the case that the spacetime includes
the origin, the metric must be regular at $r=0$ so that
\begin{equation}
C_{3-}=C_{4-}=0.
\end{equation}

We set the jump in $\beta_0$ ar $r=r_0$ to be $b_c$, and we
allow arbitrary jumps in $w_0$ and $J_{0,r}$. The jump constants,
together with the remaining unknown $C_{i\pm}$, are fixed
by requiring continuity in the metric variables as well
as satisfaction of the constraints. We find that the jump
conditions at $r=r_0$ are
\begin{equation}
\beta_{0+}=b_c, \;\; 
U_{0+}=U_{0-}, \;\;
U_{0+,r}=U_{0-,r} + \frac{2 b_c}{r_0^2},\;\;
w_{0+}=w_{0-}-2 r_0 b_c, \;\;
J_{0+}=J_{0-}, \;\;
J_{0+,r}=J_{0-,r},
\label{e-jM}
\end{equation}
with the jump in $U_{0}$ due to the term $\eth \beta_{,r}$ in the
$q^A R_{1A}$ equation, and not sourced by the matter. We have now
constructed a metric that is Ricci flat everywhere except on the
shell $r=r_0$, and we find the matter source by constructing the
Einstein tensor on the shell and using
$T^{\alpha\beta}=G^{\alpha\beta}/(8\pi)$. Every component of
$T^{\alpha\beta}$ is zero {\em except}
\begin{equation}
T^{00}=\delta(r-r_0) Z_{\ell m} b_c/(2\pi r_0).
\end{equation}
If we now set $b_c=2\pi r_0 \rho_0$, we can interpret
$T^{\alpha\beta}$ as being the stress energy tensor of a dust
shell with energy density $\rho = \delta(r-r_0) Z_{\ell m} \rho_0$.

\subsection{Case $\nu=0$, $M \ne 0$ -- Static shell on a Schwarzschild
background}
\label{s-nu0M}
\begin{eqnarray}
J_{0+} &=& 2(2M^2-r^2)C_{1+} 
+2\left(2M^2-2rM-2r^2+(2M^2-r^2)
\ln\left(1-\frac{2M}{r}\right)\right)C_{2+}
+C_{3+} +\frac{2C_{4+}}{r}+\frac{\beta_{0+}}{2}
\nonumber \\
U_{0+} &=& \frac{2\beta_{0+}}{r} -4(r-2M)C_{1+} -4 \frac{C_{2+}}{r}
\left(2r^2-2M^2-2rM+r(r-2M)\ln\left(1-\frac{2M}{r}\right)\right)
+\frac{4 C_{4+}(r+M)}{r^3} \nonumber \\
w_{0+} &=& -4 \beta_{0+} r+ 12 C_{1+} r (r-2M)^2 +12 C_{2+} r
\left(2(r^2+M^2-3rM)+(r-2M)^2\ln\left(1-\frac{2M}{r}\right)\right)
\nonumber \\
&&+12 C_{3+}r+\frac{12 M C_{4+}}{r} + C_{5+}
\end{eqnarray} 
When we substitute the above
solution into the constraints, we find that $R_{01}=0$ is satisfied
identically, and that, after removing common factors, the other
constraints reduce to
\begin{equation}
R_{00}:\;\; 
2r(\beta_{0+}-6C_{3+})+32M^3C_{2+}+8M\beta_{0+}-C_{5+}=0,\;\;\;
q^A R_{0A}:\;\; 32M^3C_{2+}+8M\beta_{0+}-C_{5+}=0.
\label{e-cS}
\end{equation}

The asymptotic form of $J_{0+}$ is
\begin{equation}
J_{0+}(r) \asymp -2(C_{1+}+2C_{2+})r^2 +4M^2(C_{1+}+2C_{2+})
+C_{3+}+\frac{\beta_{0+}}{2},
\end{equation}
so that regularity at infinity will require
\begin{equation}
C_{1+}+2C_{2+}=0,
\label{e-bi}
\end{equation}
and thus $J_{0\infty}=C_{3+}+\beta_{0+}/2$.
Using the constraints, it follows that $C_{3+}=\beta_{0+}/6$,
so that $J_{0\infty}=2\beta_{0+}/3$,
and this immediately leads to the news $N$ being zero as expected.

The jump conditions and stress-energy tensor at $r=r_0$ were
determined as in Sec.~\ref{s-M0nu0}.  The jump conditions are
\begin{eqnarray}
\beta_{0+}&=& b_c, \;\; 
U_{0+}=U_{0-}, \;\;
U_{0+,r}=U_{0-,r} + \frac{2 b_c}{r_0^2},\nonumber \\
w_{0+}&=&w_{0-}-2b_c(r_0-2M), \;\;
J_{0+}=J_{0-}, \;\;
J_{0+,r}=J_{0-,r}-\frac{ M b_c}{r_0(r_0-2M)}.
\end{eqnarray}
The only non-zero components of $T^{\alpha\beta}$ are
\begin{equation}
T^{00}=\rho\left(1-\frac{2M}{r_0}\right)^{-1},
\;\; h_{AB}T^{AB}=\rho
\frac{M}{r_0^3}\left(1-\frac{2M}{r_0}\right)^{-1},
\;\; q_A q_B T^{AB}=\eth^2\rho
\frac{M}{4r_0^3}\left(1-\frac{2M}{r_0}\right)^{-1},
\label{e-Tab}
\end{equation}
where we have set
\begin{equation}
b_c=2\pi r \rho_0 \left(1-\frac{2M}{r_0}\right)^{-1}
\mbox{ and }\rho=\delta(r-r_0) Z_{20}\rho_0,
\label{e-bcrh}
\end{equation}
so that $T^{00}=\rho (v^0)^2$ and thus $\rho$ may be interpreted as
an energy density. The above stress-energy tensor is {\em not}
that of a fluid, but represents a rigid shell in which the
stresses are such that they preserve the equilibrium of the
structure and prevent it from falling towards the black hole.
(The above form, in which the density varies with the angle,
may be unfamiliar, but the spherically symmetric situation
is more commonplace. In this case, $\rho$ in Eq.~(\ref{e-bcrh})
is replaced by
$\rho=\delta(r-r_0)\rho_0$, $T^{00}$ and $h_{AB}T^{AB}$ take
the forms given in Eq.~(\ref{e-Tab}), and $q_A q_B T^{AB}=0$;
such a stress-energy tensor describes common objects like a balloon,
or a bathysphere immersed in the sea.) Returning to the form
in Eqs.~(\ref{e-Tab}) and (\ref{e-bcrh}), we note that the stress-energy
tensor has been constructed from a metric, and therefore the Bianchi
identities imply $T^{\alpha\beta}_{;\beta}=0$; we have checked
explicitly that this identity is indeed satisfied. Since $Z_{20}$ is
negative in places, the stress-energy tensor Eq.~(\ref{e-Tab})
violates standard energy conditions~\cite{HE}. However, that difficulty
is easily rectified by supposing that the matter source is a sum of
$Z_{00}$ and $Z_{20}$ modes, with the coefficient of $Z_{00}$ larger
by a factor of $\sqrt{5}$ than that of $Z_{20}$. Then, provided
$r_0>2M$, all the standard energy conditions are satisfied.

There are 10 arbitrary constants available, and these must be chosen
so as to satisfy the following 9 conditions
\begin{itemize}
\item Regularity at the horizon $r=2M$ imposes 1 condition,
$C_{2-}=0$;
\item Applying the constraints in the exterior imposes the 2 
conditions in Eqs.~(\ref{e-cS}) -- because the jump conditions
preserve the constraints, no additional information is obtained by
imposing the constraints in the interior;
\item Regularity at infinity imposes 1 condition,
Eq.~(\ref{e-bi});
\item There are 5 jump conditions at $r=r_0$.
\end{itemize}
Thus, we construct a solution that contains 1 arbitrary constant. The
values of the various metric quantities at $r=2M$ are
\begin{equation}
U_{0-}=\frac{3C_0}{2M^2},\;\;
U_{0-,r}=-\frac{7C_0}{4 M^3}+\frac{\Phi}{M^2}, \;\;
w_{0-}= 6 C_0,\;\; J_{0-}= \frac{C_0}{M} +\Phi
\label{e-bb}
\end{equation}
where $C_0$ is the arbitrary constant, and where
\begin{equation}
\Phi=\frac{\beta_{0+}}{12M^3} \bigg(
(-3 r_0^3 +9 r_0^2 M -12 r_0 M^2 +6 M^3)
\ln\left( 1-\frac{2M}{r_0} \right) -14 M^3 +12 r_0 M^2 -6 r_0^2 M
\bigg).
\end{equation}
Thus, we may impose that both of $U_{0-}$ and $w_{0-}$, or one of
$U_{0-,r}$ or $J_{0-}$, vanish at the horizon, but we {\em cannot}
require
\begin{equation}
U=w=U_{,r}=J=0 \mbox{ at } r=2M
\label{e-bc}
\end{equation}

\subsection{Case $\nu \ne 0$, $M=0$ -- Dynamic shell on a Minkowski
background}
\label{s-nuM0}
\begin{eqnarray}
J_{0+} &=& \frac{C_{1+}}{4r}
-\frac{C_{2+}}{12r^3}
+\frac{e^{2i\nu r}C_{3+}}{4 r^3}
    \left(r^2 \nu^2 +2 i r\nu -1 \right)
+\frac{i C_{4+}}{\nu}
\nonumber \\
U_{0+} &=& \frac{C_{1+}}{2r^2}
+\frac{C_{2+}}{12r^4}\left(3+4i\nu r \right)
+\frac{e^{2i\nu r}C_{3+}}{4 r^4}
    \left(3-2i\nu r \right)
+C_{4+} + \frac{2 \beta_{0+}}{r}
\nonumber \\
w_{0+} &=&\frac{C_{2+}}{2r^2}\left(1+2i\nu r \right)
+\frac{3 e^{2i\nu r}C_{3+}}{2 r^2}
+6 r C_{4+}\left(\frac{2i}{\nu}-r \right) +C_{5+}
-10 \beta_{0+}r
\label{e-hMu}
\end{eqnarray}
When we substitute the above
solution into the constraints, we find that $R_{01}=0$ is satisfied
identically, and that, after removing common factors, the other
constraints reduce to
\begin{equation}
R_{00}:\;\; 
24r\beta_{0+}\nu+3i\nu^2rC_{1+}-3\nu^3C_{2+}-36irC_{4+}
+(ir\nu^2-3\nu)C_{5+}=0,\;\;\;
q^A R_{0A}:\;\; C_{5+}+\nu^2 C_{2+}=0.
\label{e-cMu}
\end{equation}
Regularity at null infinity leads to
\begin{equation}
C_{3+}=0.
\end{equation}
Imposing the condition that the spacetime should include
the origin means that a series expansion about $r=0$ of
Eqs.~(\ref{e-hMu}) must be regular. This leads to
\begin{equation}
C_{1-}=\nu^2 C_{3-},\;\; C_{2-}=-3 C_{3-}.
\end{equation}
The jump conditions at $r=r_0$ are
\begin{eqnarray}
\beta_{0+}&=& b_c, \;\; 
U_{0+}=U_{0-}, \;\;
U_{0+,r}=U_{0-,r} + \frac{2 b_c}{r_0^2}-\frac{4i\nu b_c}{3 r_0^2},\nonumber \\
w_{0+}&=&w_{0-}-2b_c r_0, \;\;
J_{0+}=J_{0-}, \;\;
J_{0+,r}=J_{0-,r}+\frac{ r_0 \nu^2 b_c}{3}.
\end{eqnarray}
The only non-zero components of $T^{\alpha\beta}$ are
\begin{equation}
T^{00}=f,
\;\; q_A T^{0A}=\frac{i \nu \eth f}{6},
\;\; q_A q_B T^{AB}=-\frac{\nu^2 \eth^2 f}{12},
\mbox{ where } f=\frac{b_c \delta(r-r_0) Z_{20} e^{i\nu u}}{2\pi r_0}.
\label{e-Tab1}
\end{equation}
In this case, as expected, the matter shell is dynamic with movement
of matter within the shell (because $T^{0A}\ne 0$); but we do not
investigate further the physical properties of the matter shell
Eq.~(\ref{e-Tab1}).

\subsection{Case $\nu \ne 0$, $M \ne 0$ -- Dynamic shell on a Schwarzschild
background}
\label{s-nuM}
We  were not able to obtain a solution in simple analytic form.
However, we did obtain series solutions about the singular points
of Eq.~(\ref{e-S}), i.e. about $r=2M$ and $r=\infty$.

\subsubsection{Solution about $r=2M$}
\label{s-hor}
Writing $r=y+2M$, we found that the series solution about $y=0$ of
Eq.~(\ref{e-S}) contains one singular solution and one regular
solution. We discarded the singular solution, and found
\begin{equation}
J_{0-}=-C_{3-}\frac{y^2}{8M^2} \left( 1 - 
\frac{14iM\nu-3}{3M(4iM\nu-3)} y  + \cdots \right) 
+\frac{C_{2-}}{y+2M} + C_{1-}.
\end{equation}
\begin{equation}
U_{0-}=-C_{1-}\nu i+ C_{2-}\frac{2(y+3M)}{(y+2M)^3}
-C_{3-}\frac{1-2i M\nu}{2M}\left(1-\frac{y}{M}+ \cdots \right)
\end{equation}
\begin{equation}
w_{0-}=C_{1-}6y(2+iy\nu+4iM\nu)-C_{2-}\frac{3y}{y+2M}+C_{5-}
+C_{3-}6(1-2i\nu M)y\left(1+\frac{y^2}{12M^2}+ \cdots \right)
\end{equation}

The constraints were evaluated and found to be series with
somewhat complicated coefficients, so at first sight it appeared
to be impossible to satisfy them. However, the various coefficients
are not linearly independent, and impose only two conditions on the
constants $C_{1-}$, $C_{2-}$, $C_{3-}$ and $C_{5-}$. Taking $C_{1-}$
and $C_{3-}$ as free, we found that the following solution for
$C_{5-}$ and $C_{2-}$ satisfies all the constraints at least up to
the order tested (i.e., ${\mathcal O}(y^7)$)
\begin{equation}
C_{5-}=-C_{3-}\frac{3(44iM^2\nu^2-8M\nu+5i+16M^3\nu^3)}{2\nu(2+Mi\nu)}
-C_{1-}\frac{3(8M^3\nu^3-30iM^2\nu^2-31M\nu+6i)}{\nu(2+Mi\nu)}
\end{equation}
\begin{equation}
C_{2-}=-C_{1-}\frac{3i}{\nu}
+C_{3-}\frac{-38M\nu-15i+24iM^2\nu^2+16M^3\nu^3}{6\nu(2+Mi\nu)}
\end{equation}

\subsubsection{Solution about $r=\infty$}

Here we use $r=1/x$ and the differential equation to be solved
is Eq.~(\ref{e-S}). This has an irregular singularity at $x=0$, but
nevertheless we were able to construct one regular series solution.
Using standard techniques we then
constructed the second solution for $J_2$, but found, as might be
expected, that this solution is singular at $x=0$; therefore we
discarded it. The non-singular solution is
\begin{equation}
J_{0+}= C_{1+} + C_{2+}x + C_{3+}\frac{x^3}{6} \left(1 
-\frac{9iM}{4\nu}x^2 +\cdots \right).
\end{equation}
\begin{equation}
U_{0+}=2\beta_{0+}x-C_{1+}i\nu+C_{2+}2x^2(1+Mx)
-C_{3+}x^3\frac{2i\nu}{3}\left(1-\frac{3i}{4\nu}x +\cdots \right)
\end{equation}
\begin{equation}
w_{0+}=-10\beta_{0+} x^{-1}+C_{1+}6(2x^{-1}+i\nu x^{-2}) +C_{2+}6Mx
-C_{3+}2i\nu x \left(1 - \frac{i}{2\nu}x  +\cdots \right) +C_{5+}.
\end{equation}

The situation concerning the constraints is similar to that noted in
Sec.~\ref{s-hor}. Here we take $C_{1+}$ and $C_{2+}$ as free, and find
that the following solution satisfies all the constraints up to the
order tested (i.e., ${\mathcal O}(x^9)$)
\begin{equation}
C_{5+}=-\beta_{0+}\frac{4(\nu M-6i)}{\nu}
-C_{1+}\frac{18(2i-\nu M)}{\nu} -12 C_{2+}
\end{equation}
\begin{equation}
C_{3+}=\beta_{0+}\frac{6(2i-\nu M)}{\nu^3} 
+C_{1+}\frac{9(\nu M-2i)}{\nu^3}
-C_{2+}\frac{3(2+i\nu M)}{\nu^2}
\end{equation}

\subsubsection{Comments}
\label{s-nuMc}
The number of conditions and degrees of freedom is as in the case
$\nu =0$, $M\ne 0$. Thus we expect that it is not possible to impose
$J=U=U_{,r}=w=0$ at $r=2M$, unless the solution is identically zero
throughout the interior; we have confirmed by direct calculation
that this is indeed the case.

The coefficients of the series, about both $r=2M$ and $r=\infty$,
have been calculated term by term, and general formulas for the
$n$th coefficient are not known. While the various series do appear
to have non-zero radii of convergence, it is not possible to make
any formal statement about this issue.

In order to work out the jump conditions and consequently the form
of the stress-energy tensor on the shell, we will need to match the
series solutions, about $r=2M$ and $r=\infty$, there. Most likely
this will have to be done numerically and is deferred to further
work.

\section{Conclusion}
\label{s-c}
\subsection{Discussion}
\label{s-d}

In Sec.~\ref{s-nu0M} it was shown that, under the circumstances
stated, boundary conditions Eq.~(\ref{e-bc})
may not be imposed. Further, Sec.~\ref{s-nuMc} indicates that
this result extends to the dynamic, i.e. $\nu \ne 0$, case.
Since the imposition of such boundary conditions is common
practice~\cite{hpn,mat,particle,mod}, the implications
of the result need further discussion.

First, let us be clear about the conditions under which the
result applies.
\begin{enumerate}
\item There needs to be a matter source that is not
spherically symmetric around a Schwarzschild black hole. The
matter source must be either static, or undergoing regular
oscillations ($\propto e^{i\nu u}$). 
\item All the metric coefficients need to be either static or
undergoing regular oscillations.
\end{enumerate}
Let us note here that the matter
source is crucial to the argument, because a vacuum perturbation
of Schwarzschild cannot be static or be undergoing regular
oscillations: the ``no-hair'' theorem indicates that such a
spacetime must relax to pure Schwarzschild (or, in general, to
Kerr if the initial perturbation contains angular momentum).

\begin{figure}
\begin{center}
\epsfig{file=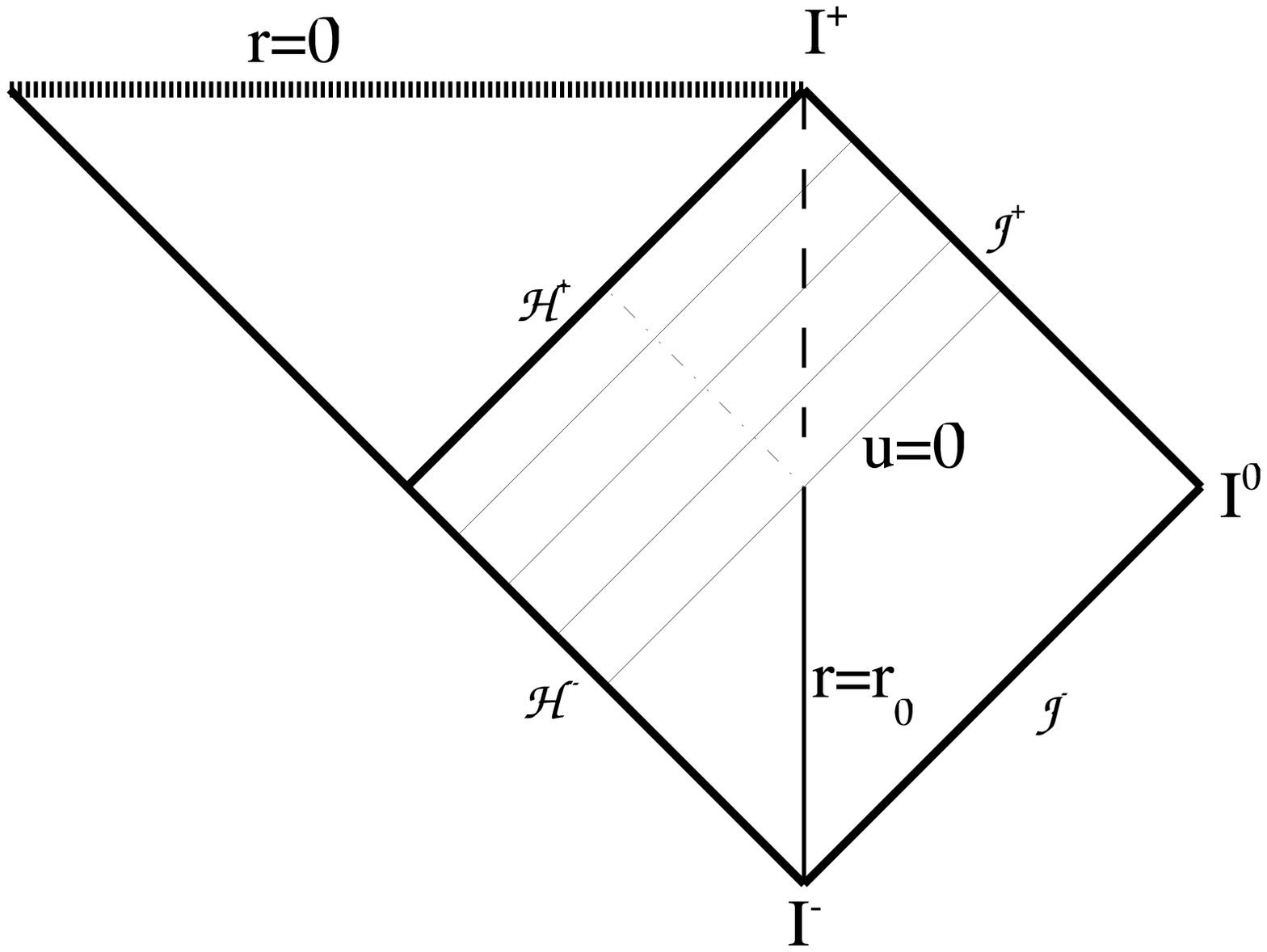,height=8in}
\end{center}
\caption{The Penrose diagram of the spacetime described in Sec.~\ref{s-d}.
${\mathcal H}^-$ and ${\mathcal H}^+$ are the past and future horizons of
the Schwarzschild black hole. The matter shell at $r=r_0$ is the straight
line between $I^-$ and $I^+$. The line becomes dotted at $u=0$ indicating
a change from spherical symmetry, and the thin dotted line going to
${\mathcal H}^+$ indicates the inner boundary of the domain of influence
of this change.}
\label{f-p}
\end{figure}

We investigate further by supposing that condition 1 above applies
but not condition 2, and for simplicity we consider the static
case. We make the {\em gedenken} experiment that a spherically
symmetric shell has existed around a black hole since
$u=-\infty$. At $u=0$ the matter starts to move and a new
static configuration, with $\rho$ a sum of $Z_{00}$ and
$Z_{20}$ modes, is in place from $u=M$ until $u=+\infty$.
The situation is illustrated in Fig.~\ref{f-p}.
For $u < 0$, $\beta=U=U_{,r}=w=J=0$ at the boundary as well as
everywhere inside the shell. For $u>0$, changes to the
gravitational field propagate away from the shell at the speed
of light, and the metric relaxes towards a form that satisfies
Eq.~(\ref{e-bb}) at $r=2M$.
The inner boundary of the coordinate system is $r=2M$ on
the past event horizon ${\mathcal H}^-$ and is an ingoing null
surface. Thus the effects of the shell changing cannot reach it,
and the metric $g_{\alpha\beta}$ is well-behaved on each null
cone $u=$constant. However, 
\[\lim_{u\rightarrow \infty}g_{\alpha\beta}\]
is not continuous at $r=2M$. Thus we conclude
that if only condition 1 above applies, then homogeneous
boundary conditions Eq.~(\ref{e-bc}) may be applied for a finite
time lapse $u$, but not for an infinite or semi-infinite time.

In a numerical simulation using homogeneous boundary conditions
Eq.~(\ref{e-bc}) and condition 1,
the relaxation of the metric towards the boundary conditions
Eq.~(\ref{e-bb}), means that
the metric variables must make a finite jump between, say,
$r=2M$ and $r=2M+\Delta r$.
Now, $\Delta r \rightarrow 0$ as $u \rightarrow \infty$, so for
short evolutions there is no problem with homogeneous boundary
condions; but for longer evolutions, once
$\Delta r$ becomes smaller than the grid discretization, the
solution will seem to be discontinuous and inaccuracies will
develop thus making code predictions unreliable. 

The difficulty with imposing appropriate boundary data at a
black hole horizon in the Bondi-Sachs formalism does not, to
our knowledge, affect any results already in the literature.
As already stated, the problem arises only if
matter is present, and so for example it does not apply to
computations involving the scattering of gravitational
radiation by a Schwarzschild black hole~\cite{mod}. Further, the
effect comes into play only as $u \rightarrow \infty$, and so
does not apply to~\cite{mat} where the computations
terminate at $u={\mathcal O}(M)$. The earlier work on particle
evolution~\cite{particle} does have long term evolutions, but
these were not presented as physical results.

\subsection{Opportunities for further work}
As described in Sec.~\ref{s-in}, the results obtained here are
expected to be useful in the validation and improvement of
numerical codes using the Bondi-Sachs formalism, particularly
in the computation of gravitational radiation. On this point, we note
that the coefficient of $J_{0\infty}$ in Eq.~(\ref{e-N}) scales,
for large $\ell$, as $\ell^4$, and so it may be that noise at
high angular frequency is being multiplied by a large number.
Further investigation and numerical testing is required, the starting
point for which would probably be the solution exterior
to the shell of Sec.~\ref{s-nuM0} ($\nu\ne 0,M=0$).

This paper has raised the question as to the appropriate
boundary data at the horizon for a physical situation involving
matter near a Schwarzschild black hole. However, except for the
specific case of a static distribution of matter in the form of
a thin shell, the issue has not been resolved. Clearly, a
prescription is needed when there is a general, dynamic matter
distribution.

It is in principle possible to develop the solution obtained in
Sec.~\ref{s-nuM} ($\nu\ne 0,M\ne 0$) to yield
the gravitational radiation emitted by a star in close orbit
around a black hole, because in this case the metric variables
can be expressed as a sum of terms of the form Eq.~(\ref{e-an}).
While this has already been accomplished within a different
formalism~\cite{glamp,sah}, an alternative derivation would
nevertheless be useful. As noted in Sec.~\ref{s-nuM0}, we have
not investigated in any detail the physical properties of a
time-varying matter shell; it would be necessary to do so in
order to pursue this project. 

Evaluation of the constraints ($R_{0\alpha}=0$ in vacuum) has
been an essential feature of this paper. Numerical codes based
on the characteristic formalism use the constraints in setting
boundary data on the inner worldtube -- for
example homogeneous boundary data (Eq.~\ref{e-bc})) trivially
satisfy the constraints, and the constraints have been
used~\cite{fission}, in a dynamic situation, to determine part of
the data. In ADM type codes it is common practice to evaluate the
constraints in the interior of the spacetime in order to monitor
the accuracy of the evolution. Characteristic codes do not
normally evaluate the constraints off the boundary (an exception
being~\cite{bartnik}), but perhaps this could lead to useful
insights.

\section*{Acknowledgments}
I wish to thank Shrirang Deshingkar, Luis Lehner and Jeffrey
Winicour for discussions.
I thank Horace Hearne Laboratory of Theoretical Physics, Louisiana State
University, as well as Max-Planck-Institut f\"ur
GravitationsPhysik, Albert-Einstein-Instit\"ut, for hospitality. 
The work was supported in part by the National Research Foundation,
South Africa, under Grant number 2053724, and by the NSF under grant
NSF-INT 0242507 to Louisiana State University.

\end{document}